\documentclass{article}
\usepackage{amsfonts}

\catcode`\@=11 \@addtoreset{equation}{section}\catcode`\@=12

\newcommand{\beq}{\begin{equation}}
\newcommand{\eeq}{\end{equation}}
\newcommand{\eeqn}{\nonumber \end{equation}}
\newcommand{\bear}{\begin{eqnarray} &&\nqq}
\newcommand{\bearr}{\begin{eqnarray}}
\newcommand{\ear}{\end{eqnarray}}
\newcommand{\earn}{\nonumber \end{eqnarray}}

\newcommand{\nn}{\nonumber\\ &&\nqq}
\newcommand{\mm}{\\ &&\nqq}

\newcommand{\nqq}{\hspace{-2em}}

\newcommand{\R}{{\mathbb R}}

\newcommand{\diag}{\mathop{\rm diag}\nolimits}

\newcommand{\const}{\mathop{\rm const}\nolimits}

\newcommand{\p}{\partial}

\newcommand{\fnm}{\footnotemark}
\newcommand{\fnt}{\footnotetext}

\newcommand{\aver}[1]{\langle\,#1\,\rangle\mathstrut}
\newcommand{\noi}{\noindent}

\newcommand{\tst}{\textstyle}

\newcommand{\fract}[2]{{\tst\frac{#1}{#2}}}

\newcommand{\half}{{\fract{1}{2}}}

\newcommand{\mn}{_{\mu\nu}}
\newcommand{\MN}{^{\mu\nu}}
\newcommand{\mN}{_\mu^\nu}

\newcommand{\cR}{{\cal R}}
\newcommand{\ten}[1]{\mbox{$\cdot 10^{#1}$}}

\begin{document}

\thispagestyle{empty}

\vspace{15pt}

 \begin{center}
 \Large\bf
 MODELS OF \emph{G} TIME VARIATIONS IN DIVERSE DIMENSIONS \\[15pt]
 \large
 V.N. Melnikov\fnm[1]\fnt[1]{melnikov@phys.msu.ru, melnikov@vniims.ru } \\[10pt]

 \vspace{10pt}

 \normalsize\it
 Center for Gravitation and Fundamental Metrology, VNIIMS, \\
 46 Ozyornaya Str., Moscow 119361, Russia and \\[5pt]
 Institute of Gravitation and Cosmology, People's Friendship University \\
 of Russia, 6 Mikhlukho-Maklaya Str., Moscow 117198, Russia
 \end{center}

 \vspace{15pt}

 \begin{abstract}

A review of different cosmological models in diverse dimensions
leading to a relatively small time variation of the effective
gravitational constant $G$ is presented. Among them: 4-dimensional
general scalar-tensor model, multidimensional vacuum model with
two curved Einstein spaces, multidimensional model with
multicomponent anisotropic ``perfect fluid'', $S$-brane model with
scalar fields and two form field etc. It is shown,  that there
exist different possible ways of explanation of relatively small
time variation of   the effective gravitational constant $G$
compatible with present cosmological data (e.g. acceleration):
4-dimensional scalar-tensor theories or multidimensional
cosmological models with different matter sources. The
experimental bounds on $\dot {G}$ may be satisfied ether in some
restricted interval or for all allowed values of the synchronous
time variable.

\end{abstract}

\hspace{1cm} PACS: 04.50.-h, 04.80.Cc, 06.20.Jr, 95.36.+x

\hspace{1cm} Keywords: multidimensional cosmological models,
scalar-tensor theories, gravitational constant, acceleration of
the Universe, dark energy, variations of $G$

\newpage

\section{Introduction}

In the development of relativistic gravitation and dynamical
cosmology after A. Einstein and A. Friedmann there are three
distinct stages: first, investigation of models with matter
sources in the form of a perfect fluid, as was originally done by
Einstein and Friedmann. Second, studies of models with sources as
different physical fields, starting from electromagnetic and
scalar ones, both in classical and quantum cases (see
\cite{Stan}). And third, which is really topical now, application
of ideas and results of unified models for treating fundamental
problems of cosmology and black hole physics, especially in high
energy regimes and for explanation of the greatest challenge to
modern physics - the present acceleration of the Universe, dark
matter and dark energy problems. Multidimensional gravitational
models play an essential role in the latter approach.

The necessity of studying multidimensional models of gravitation
and cosmology \cite{Mel2,Mel,Mel3} is motivated by several
reasons. First, the main trend of modern physics is the
unification of all known fundamental physical interactions:
electromagnetic, weak, strong and gravitational ones. During the
recent decades there has been a significant progress in unifying
weak and electromagnetic interactions, some more modest
achievements in GUT \cite{KMKh}, super-symmetric, string and
super-string theories.

Now, theories with membranes, $p$-branes and more vague M-theory
are being created and studied. Having no definite successful
theory of unification now, it is desirable to study common
features of these theories and their applications to solving basic
problems of gravitation and cosmology. If we really believe in
unified theories, the early stages of the Universe evolution and
black hole physics, as unique superhigh energy regions and
possibly even low energy stage, where we observe the present
acceleration, are the most proper and natural arena for them.

Second, multidimensional gravitational models, as well as
scalar-tensor theories of gravity, are theoretical frameworks for
describing possible temporal and range variations of fundamental
physical constants \cite{Stan,4,5,6}. These ideas have originated
from papers of E. Milne (1935) and P. Dirac (1937) on relations
between the phenomena of micro- and macro-worlds, and up till now
they are under thorough study both theoretically and
experimentally. The possible discoveries of the fine structure
constant and proton to electron ratio variations are now at a
critical further investigation.

Lastly, applying multidimensional gravitational models to basic
problems of modern cosmology and black hole physics, we hope to
find answers to such long-standing problems as singular or
nonsingular initial states, creation of the Universe, creation of
matter and its entropy, cosmological constant, coincidence
problem, origin of inflation and specific scalar fields which may
be necessary for its realization, isotropization and graceful exit
problems, stability and nature of fundamental constants
\cite{4,Solv,Paris,KM}, possible number of extra dimensions, their
stable compactification, new revolutionary data on  present
acceleration of the Universe, dark matter and dark energy
\cite{Paris} etc.

Bearing in mind that multidimensional gravitational models are
certain generalizations of general relativity which is tested
reliably for weak fields up to 0.0001 and partially in strong
fields (binary pulsars), it is quite natural to inquire about
their possible observational or experimental windows.  From what
we already know, among these windows are \cite{Solv,Paris}:

-- possible deviations from the Newton and Coulomb laws, or new
interactions,

-- possible variations of the effective gravitational constant
with a time rate smaller than the Hubble one,

-- possible existence of monopole modes in gravitational waves,

-- different behavior of strong field objects, such as
multidimensional black holes, wormholes and $p$-branes,

-- standard cosmological tests,

-- possible non-conservation of energy in strong field objects and
accelerators, if brane-world ideas about gravity in the bulk turn
out to be true etc.

Since modern cosmology has already become a unique laboratory for
testing standard unified models of physical interactions at
energies that are far beyond the level of existing and future
man-made accelerators and other installations on the Earth, there
exists a possibility of using cosmological and astrophysical data
for discriminating between future unified schemes. Data on
possible time variations or possible deviations from the Newton
law as a new important test should also contribute to the unified
theory choice and viable cosmological model choice as well
\cite{PIRT-07,Mtest}.

As no accepted unified model exists, in our approach
\cite{Mel2,Mel,Mel3,IMtop} we adopted simple (but general from the
point of view of number of dimensions) models, based on
multidimensional Einstein equations with or without sources of
different nature:

-- cosmological constant,

-- perfect and viscous fluids,

-- scalar and electromagnetic fields,

-- their possible interactions,

-- dilaton and moduli fields with or without potentials,

-- fields of antisymmetric forms (related to $p$-branes) etc.

Our program's main objective was and is to obtain exact
self-consistent solutions (integrable models) for these models and
then to analyze them in cosmological, spherically and axially
symmetric cases. In our view this is a natural and most reliable
way to study highly nonlinear systems.  It is done mainly within
Riemannian geometry. Some simple models in integrable Weyl
geometry and with torsion were studied as well. In many cases we
tried to single out models, which do not contradict available
experimental or observational data on variations of $G$. In some
cases we used our methods for arbitrary dimensions in studying 4D
models also.

As our model \cite{Mel2,Mel,Mel3} we use n Einstein spaces of
constant curvature with sources as (m+1)-component perfect fluid,
(or fields or form-fields,…), cosmological or spherically
symmetric metric, manifold as a direct product of factor-spaces of
arbitrary dimensions. Then, in harmonic time guage  we show that
Einstein multidimensional equations are equivalent to Lagrange
equations with non-diagonal in general mini-super-space metric and
some exponential potential. After diagonalization of this metric
we perform reduction to sigma-model \cite{IM12,IMC} and Toda-like
systems \cite{GIM1}, further to Liouville, Abel, generalized
Emden-Fowler Eqs. etc. and try to find exact solutions. We suppose
that behavior of extra spaces is the following: they are constant,
or dynamically compactified, or like torus, or large, but with
barriers, walls etc.

So, we realized and continue to realize the program in arbitrary
dimensions (from 1988) \cite{Mel2,Mel,Mel3,IMtop,Paris} :

In cosmology we obtained exact general solutions of
multidimensional Einstein equations with sources:

 - $\Lambda$ ,  $\Lambda$ + scalar field (e.g. nonsingular, dynamically compactified, inflationary) \cite{BIMZ};

 - perfect fluid, PF + scalar field (e.g. nonsingular, inflationary solutions) \cite{IM5,IM10};

 - viscous fluid (e.g. nonsingular, generation of mass and entropy, quintessence and
   coincidence in 2-component model);

 - stochastic behavior near the singularity, billiards in Lobachevsky space,
D=11 is critical, $\varphi$ destroys billiards (1994).

(For all above cases Ricci-flat solutions were obtained for any n)

- also solutions with curvature in one factor-space;

- with curvatures in 2 factor-spaces only for total N=10, 11;

 - with fields: scalar, dilatons, forms of arbitrary rank \cite{IM11} (1998) -
inflationary, $\Lambda$ generation by forms (p-branes)
 \cite{Cosm};

 - first billiards for dilaton-forms (p-branes) interaction (1999);

 - quantum variants
(solutions of WDW-equation \cite{BIMZ,IM10,IMZ}) for all above
cases where classical solutions were obtained;

 - dilatonic fields with potentials, billiard behavior for them.

 For many of these integrable models we calculated also the
 variation with time of the effective
 gravitational constant. Comparison with present experimental
 bounds allowed to choose particular models or single out some
 classes of solutions.

 Similar methods were used for obtaining
 exact solutions in spherical symmetry case \cite{Mel2,Mel,Mel3,IMtop,Paris}.

Dirac's Large Numbers Hypothesis (LNH) is the origin of many
theoretical studies of time-varying $G$. According to LNH, the
value of $\dot{G}/G$ should be approximately the Hubble rate.
Although it has become clear in recent decades that the Hubble
rate is too high to be compatible with experiments, the enduring
legacy of Dirac's bold stroke is the acceptance by modern theories
of non-zero values of $\dot{G}/G$ as being potentially consistent
with physical reality.

There are three problems related to $G$, whose origin lies mainly
in predictions of unified models of physical interactions:

1) absolute $G$ measurements, 2) possible time variations of $G$,
3) possible range variations of $G$ -- non-Newtonian, or new
interactions.

After the original {Dirac hypothesis} some new ones appeared and also some
generalized {theories} of gravitation admitting variations of the effective
gravitational coupling. We can single out three stages in the development of
this field \cite{4}:

1. Study of phenomenological theories and hypotheses with
variations of FPC, their predictions and confrontation with
experiments (1937-1977).

2.Creation of theories admitting variations of an effective
gravitational constant in a particular system of units, analyses
of experimental and observational data within these theories
\cite{Stan} (1977-present).

3. Analysis of FPC variations within unified models \cite{Mel}
(present).

Different theoretical schemes lead to temporal variations of the
effective gravitational constant \cite{mel76}:

- Empirical models and theories of Dirac type, where $G$ is
replaced by $G(t)$.

- Numerous scalar-tensor theories of Jordan-Brans-Dicke type where
$G$ depending on the scalar field $\phi(t)$ appears.

- Gravitational theories with a conformal scalar field arising in
different appro\-a\-ches \cite{BMS-ITP,Stan} (they can be treated
as special cases of scalar-tensor theories).

- Multidimensional unified theories in which there are dilaton
fields and effective scalar fields appearing in our 4-dimensional
spacetime from additional dimensions \cite{Mel}. They may help
also in solving the problem of the variable cosmological constant
(from Planckian to present values) and the cosmic coincidence
problem.

A striking feature of the present status of theoretical physics is
that there is no satisfactory theory unifying all four known
interactions; most modern unification theories do not admit unique
and universal constant values of physical constants and of the
Newtonian gravitational coupling constant $G$ in particular.
Although the bounds on $\dot G$ and
 $G(r)$ are in some classes of theories rather wide on purely
theoretical grounds since any theoretical model contains a number
of adjustable parameters, we note that observational data
concerning other phenomena, in particular cosmological data, may
place limits on possible ranges of these adjustable parameters.
But, in any case variations of \emph{G} may be an additional test
of unified models, generalized theories of gravitation and
cosmological models as well \cite{Mtest}.

Here we restrict ourselves to the problem of $\dot G$ (for $G(r)$ see
\cite{Mel,4,Stan,5}). We show that various theories predict the value of $\dot{G}/G$
to be $10^{-12}/$yr or less. The significance of this fact for experimental
and observational determinations of the value of or upper bound on $\dot G$ is
the following: any determination with error bounds significantly below
 $10^{-12}/$yr (combined with experimental bounds on other parameters) will
typically be compatible with only a small portion of existing
theoretical models and will therefore cast serious doubts on the
viability of all other models. In short, a tight bound on $\dot
G$, in conjunction with other astrophysical observations, will be
a very effective ``theory killer" and/or significantly reduce the
class of viable theories. Any step forward in this direction will
be of utmost significance \cite{Mor,M-G,Taiwan}.

Some estimations for $\dot G$ were done long ago in the frames of
general scalar tensor theories using the values of cosmological
parameters ($\Omega$, $H$, $q$ etc) known at that time
\cite{ZM,IM1,BIM1,Melgc,Mel,Stan}. It is easy to show that for
modern values they predict $\dot{G}/G$ at the level of
$10^{-13}/$yr and less (see also estimations of A. Miyazaki
\cite{Mi}, predicting time variations of $G$ at the level of
$10^{-13} {\rm yr}^{-1}$ for the Machian-type cosmological
solution in the Brans-Dicke theory, Y. Fujji (see in \cite{Fj}),
of J.P.Mbelek and M.Lanchieze-Ray \cite{ML2} on the level of
$10^{-17}$/yr for the simple 5D KK-theory with an external scalar
field and section 2 of the present paper).

The most reliable experimental bounds on $\dot{G}/G$ (radar
ranging of spacecraft and planets dynamics \cite{Hel,P-97} and
laser lunar ranging \cite{Dic,N-03}) give the limit less than
$10^{-12}/$yr, so any result at less than this level will be very
important for solving the fundamental problem of variations of
constants and for discriminating between viable unified theories
and cosmological models. So, further data on lunar laser and radar
ranging and realization of MICROSCOPE, ASTROD, LATOR space
projects and such multipurpose new generation type space
experiment as Satellite Energy Exchange (SEE) for measuring $\dot
G$, absolute value of $G$ and Yukawa type forces at meters and
Earth radius ranges \cite{san} become extremely topical.

In what follows, we shall discuss predictions for $\dot G$ from
 generalized scalar-tensor theories and some multidimensional
models.

\section{Scalar-Tensor Cosmology and Variations of $G$}

The purpose of this section is to estimate the order of magnitude
of the gravitational constant $G$ variations due to cosmological
expansion in the framework of general scalar-tensor theories (STT)
of gravity \cite{BMN}.

Consider the general (Bermann-Wagoner-Nordtvedt) class of STT
where gravity is characterized by a metric $g\mn$ and a scalar
field $\phi$; the action is \beq \label{act} S = \int d^4 x
\sqrt{g}\{ f(\phi) \cR[g] + h(\phi)g\MN\phi_{,\mu}\phi_{,\nu} -2
U(\phi) + L_m \}. \eeq Here $\cR[g]$ is the scalar curvature, $g =
|\det (g_{\mu \nu})|$; $f,\ h$ and $U$ are certain functions of
$\phi$, varying from theory to theory, $L_m$ is a matter
Lagrangian.

This formulation of the theory corresponds to the Jordan conformal
frame, in which matter particles move along geodesics and hence
the weak equivalence principle is valid and non-gravitational
fundamental constants do not change. In other words, this is the
frame well describing the existing laboratory, geophysical and
cosmological observations.

Among the three functions of $\phi$ entering into (\ref{act}) only
two are independent since there is a freedom of transformations
$\phi = \phi(\phi_{\rm new})$ \cite{ZM}. We use this
arbitrariness, choosing $h(\phi) \equiv 1$, as is done, e.g., in
\cite{star00}. Another standard parametrization is to put
$f(\phi)=\phi$ and $h(\phi) = \omega(\phi)/\phi$ (the Brans-Dicke
parametrization of the general theory (\ref{act})) In our
parametrization $h\equiv 1$, the B-D parameter $\omega(\phi) = f
(f_\phi)^{-2}$; the subscript $\phi$ denotes a derivative with
respect to $\phi$. The B-D STT is the particular case $\omega =
\const$, so that in (\ref{act}) \beq \label{BD} f(\phi) =
\phi^2/(4\omega), \quad h \equiv 1. \eeq For the conformal scalar
field case see \cite{Stan,ZM}.

The field equations that follow from (\ref{act}) read
\bear \label{Ephi}
\Box \phi - \half \, \cR \, f_\phi + U_\phi = 0, \mm
f(\phi) \Bigl(\cR\mN - \half \delta\mN \cR\Bigr) = -\phi_{,\mu} \phi^{,\nu}
+ \half\delta\mN\phi^{,\alpha}\phi_{,\alpha} - \delta\mN U(\phi) 
+ (\nabla_{\mu}\nabla^\nu- \delta\mN \Box)f \label{EE} - T\mN{}_{\rm (m)},
\ear
where $\Box$ is the D'Alembert operator, and the last term in (\ref{EE}) is the
energy-momentum tensor of matter.

Consider now isotropic cosmological models with the standard FRW
metric \beq \label{ds} ds^2 = dt^2
-a^2(t)\biggl[\frac{dr^2}{1-kr^2} + r^2 (d\theta^2 +
\sin^2\theta\,d\varphi^2)\biggr], \eeq where $a(t)$ is the scale
factor of the Universe, and $k=1,\ 0,\ -1$ for closed, spatially
flat and hyperbolic models, respectively. Accordingly, we assume
$\phi = \phi(t)$ and the energy-momentum tensor of matter in the
perfect fluid form $T\mN{}_{\rm (m)} = \diag(\rho, -p, -p, -p)$
($\rho$ is a density and $p$ is a pressure).

The field equations in this case can be written as follows: \bear
\label{E0} \ddot\phi + 3\frac{\dot a}{a} \dot{\phi}
\label{Ef}-\frac{3}{a^2} (a\ddot{a} + \dot{a}{}^2 +k) + U_{\phi}
=0, \mm \frac{3f}{a^2}(\dot{a}{}^2+k)= \half \dot\phi{}^2 +U
-3\frac{\dot a}{a} \dot f +\rho, \mm
\frac{f}{a^2}(2a\ddot a +
\dot{a}{}^2 +k) \label{E1}= -\half \dot\phi{}^2 +U- \ddot
f-2\frac{\dot a}{a} \dot f -p. \ear

To connect these equations with observations, let us fix the time
$t$ at the present epoch (i.e., consider the instantaneous values
of all quantities) and introduce the standard observables: $H=\dot
a/a$ (the Hubble parameter), $q=-a\ddot a/\dot a{}^2$ (the
deceleration parameter), $\Omega_m = \rho/\rho_{\rm cr}$ (the
matter density parameter), where $\rho_{\rm cr}$ is the critical
density, or, in our model, the r.h.s. of equation (\ref{E0}) in
case $k=0$: $\rho_{\rm cr} = 3fH^2$. This is slightly different
from the usual definition $\rho_{\rm cr} = 3H^2/8\pi G$. The point
is that the locally measured Newtonian constant in STT differs
from $1/(8\pi f)$; provided the derivatives $U_{\phi\phi}$ and
$f_{\phi\phi}$ are sufficiently small, one has \cite{star00} \beq
\label{Geff} 8\pi G_{\rm eff} = \frac{1}{f} \frac{2\omega
+4}{2\omega+3}. \eeq
Since, according to the Solar-system
experiments, $\omega\geq 40000$, for our order-of-magnitude
reasoning we can safely put $8\pi G = 1/f$, and, in particular,
our definition of $\rho_{\rm cr}$ now coincides with the standard
one.

The time variation of $G$, to a good approximation, is \beq
\label{Gdot} \dot G/G \approx - \dot f/f = gH, \eeq where, for
convenience, we have introduced the coefficient $g$ expressing
$\dot G/G$ in terms of $H$.

Equations (\ref{Ef})--(\ref{E1}) contain too many arbitrary parameters for
making a good estimate of $g$. Let us now introduce some restrictions according
to the current state of observational cosmology:

\medskip\noi
{\bf (i)} $k=0$ (a spatially flat cosmological model, so that the total density
of matter equals $\rho_{\rm cr}$);

\medskip\noi
{\bf (ii)} $p=0$ (the pressure of ordinary matter is negligible compared to the
energy density);

\medskip\noi
{\bf (iii)} $\rho = 0.3\,\rho_{\rm cr}$ (the ordinary matter, including its
dark component, contributes to only 0.3 of the critical density; unusual
matter, which is here represented by the scalar field, comprises the remaining
70 per cent).

Then equations (\ref{E0}) and (\ref{E1}) can be rewritten in the
form \beq \label{E0'} \frac{1}{2}\dot\phi{}^2 + U - 3H\dot f = 2.1
H^2 f, \quad -\frac{1}{2}\dot\phi{}^2 + U - 2H\dot f -\ddot f =
(1-2q) H^2 f. \eeq Subtracting second from the first one, we
exclude the ``cosmological constant'' $U$ and obtain \beq
\label{*} \dot\phi{}^2 - H\dot f + \ddot f = (1.1 + 2q) H^2 f.
\eeq

The first term in equation (\ref{*}) can be represented in the
form \beq \dot\phi{}^2 = \dot f{}^2 (df/d\phi)^{-2} = \dot f{}^2
\omega/f, \eeqn and $\dot f/f$ can be replaced with $-gH$. The
term $\ddot f$ can be neglected for our estimation purposes for an
arbitrary function $f$ and potential $U(\phi)$.

Then, (\ref{*}) divided by $H^2 f$ leads to the quadratic equation
with respect to $g$: \beq \omega g^2 +g - q' =0 ,   q' = 1.1 +2q.
\label{**} \eeq
 According to present observations, the Universe
is expanding with an acceleration, so that the parameter $q$ is,
roughly, $-0.5\pm 0.2$, hence we can take $|q'| \leq 0.4$.

In case $q'=0$ we simply obtain $g = -1/\omega$. Assuming \beq H =
h_{100}\cdot 100\ {\rm km}/({\rm s.Mpc})\approx h_{100} \cdot
10^{-10} \ {\rm yr}^{-1} \eeqn and present limit $\omega\geq
40000$, we come to the estimate \beq \label{est1} |\dot G/G| \leq
4 \ten{-15} h_{100}\ {\rm yr}^{-1}, \eeq where $h_{100}$ is, by
modern views, close to 0.7. So (\ref{est1}) becomes \beq |\dot
G/G| \leq 4 \ten{-15}\ {\rm yr}^{-1}. \label{est1'} \eeq

For nonzero values of $q'$, solving the quadratic equation
(\ref{**}) and assuming $q'\omega\gg 1$, we arrive at the estimate
$|g| \sim \sqrt{q'/\omega}$, so that, taking $q'=0.4$ and again
$\omega\geq 40000$, we have instead of (\ref{est1}) \beq |\dot
G/G| \leq 0.9 \ten{-13} h_{100}\ {\rm yr}^{-1} \approx
0.7\ten{-13} {\rm yr}^{-1}, \label{est2} \eeq where we have again
put $h_{100}=0.7$.

As a result, in the framework of the general STT, present
cosmological observations, taking into account the Solar-system
data, restrict the possible variation of $G$ to values less then
$10^{-13}$/yr. This estimate may be considerably tightened if the
matter density parameter $\Omega_m$ and the (negative)
deceleration parameter $q$ will be determined more precisely.

\section{G-dot in $(1+ 3+N)$-dimensional cosmology with multicomponent
anisotropic fluid }

We consider here a $(4+N)$-dimensional cosmology with an isotropic
3-space and an Einstein internal space \cite{BIM1,IMW}. The
Einstein equations provide a relation between ${\dot{G}/G}$ and
other cosmological parameters.

\subsection{The model}

Let us consider $(4+N)$-dimensional theory with the gravitational part of the
action
\beq
S_g = \frac{1}{2\kappa^2}\int d^{4+N}x\sqrt{-g}R\ ,
\eeq
where $\kappa^2$ is the fundamental gravitational constant. Then the
gravitational field equations are
\beq
R^M_P = \kappa^2(T^M_P-\delta^M_P\frac{T}{N+2})\ ,
\eeq
where $T^M_P$ is a $(4+N)$-dimensional energy-momentum tensor, $T=T^M_M$, and
$M,P=0,\dots,N+3$.

For the $(4+N)$-dimensional manifold we assume the structure
\beq
M^{4+N} = \R_{*} \times M^3_k\times K^N\,
\eeq
where $\R_{*}$ is 1-dimensional time manifold, $M^3_k$ is a 3-dimensional space
of constant curvature, $M^3_k=S^3,\ R^3,\ L^3$ for $k=+1,0,-1$, respectively,
and $K^N$ is a $N$-dimensional Einstein manifold.

The metric is taken in the form
\beq
ds^2 = g_{MN}dx^Mdx^N = - dt^2 + a^2(t)g^{(3)}_{ij}(x)dx^idx^j +
b^2(t)g^{(N)}_{mn}(y)dy^mdy^n \ ,
\eeq
where $i,j,k=1,2,3;\ m,n,p=4,\dots,N+3;\ g^{(3)}_{ij},\ g^{(N)}_{mn},\ a(t)$ and
$b(t)$ are, respectively, the metrics and scale factors for $M^3_k$ and $K^N$.

For $T^M_P$ we adopt the expression of the multicomponent (anisotropic) fluid
form
\beq
(T^M_P) = \sum_{\alpha = 1}^m{\rm diag}(- \rho^{\alpha} (t), p_3^{\alpha}(t)
\delta^i_j,p_N^{\alpha}(t)\delta^m_n).
\eeq

Under these assumptions the Einstein equations take the form
\bear \label{6}
\frac{3\ddot{a}}{a} + \frac{N \ddot{b}}{b} =\frac{\kappa^2}{N+2} \sum_{\alpha = 1}^m
[-(N+1)\rho^{\alpha}-3p_3^{\alpha} -N p_N^{\alpha}], \mm
\label{7}
\frac{\ddot{a}}{a} + \frac{2 \dot{a}^2}{a^2} + \frac{N \dot{a} \dot{b}}{ab} +
\frac{2k}{a^2} = \frac{\kappa^2}{N+2} \sum_{\alpha = 1}^m [\rho^{\alpha} +
(N-1) p_3^{\alpha} - N p_N^{\alpha}], \mm
\label{8}
\frac{\ddot{b}}{b} + (N-1) \frac{\dot{b}^2}{b^2} + \frac{3 \dot{a} \dot{b}}{ab} +
\frac{\lambda}{b^2} = \frac{\kappa^2}{N+2} \sum_{\alpha = 1}^m[\rho^{\alpha} -
3 p_3^{\alpha}+ 2 p_N^{\alpha}].
\ear

Here \beq \label{8a} R_{m n}[g^{(N)}] = \lambda g^{(N)}_{mn}, \eeq
$m,n = 1,\ldots, N$, where $\lambda$ is constant. The
4-dimensional density is \beq \label{9} \rho^{\alpha,(4)}(t) =
\int_K d^Ny\sqrt{g^{(N)}}b^N(t)\rho^{\alpha}(t) =
\rho^{\alpha}(t)b(t), \eeq where we have normalized the factor
$b(t)$ by putting \beq \label{10} \int_K d^Ny\sqrt{g^{(N)}} = 1.
\eeq

On the other hand, to get the 4-dimensional gravity equations one should put
$8\pi G(t)\rho^{\alpha(4)}(t) = \kappa^2\rho^{\alpha}(t)$.

Consequently, the effective 4-dimensional gravitational
``constant'' $G(t)$ is defined by \beq 8\pi G(t) =
\kappa^2b^{-N}(t),
 \eeqn whence its time variation is expressed as
\beq \label{G-dot} \dot{G}/G = -N \dot{b}/b. \eeq

\subsection{Cosmological parameters}

Some inferences concerning the observational cosmological parameters can be
extracted just from the equations without solving them \cite{BIM1}. Indeed, let
us define the Hubble parameter $H$, the density parameters $\Omega^{\alpha}$
and the "deceleration" parameter $q$ referring to a fixed instant $t_0$ in the
usual way
\beq \label{cosm}
H = \dot{a}/a,  \qquad \Omega^{\alpha} = 8\pi G\rho^{\alpha,(4)}/3H^2 =
\kappa^2\rho^{\alpha}/3H^2, \quad  q = -a  \ddot{a}/\dot{a}^2\ .
\eeq

Besides, instead of $G$ let us introduce the dimensionless parameter
\beq \label{g-dot}
g = \dot{G}/GH = -Na \dot{b}/\dot{a}b .
\eeq

The present observational upper bound is  g < 0.1, if we take in
accord with  \cite{Hel,Dic}
\beq \label{18b} \dot{G}/G  <
 10^{-12}(y^{-1}) \eeq
 and $H = (0.7 \pm 0.1) \times
10^{-11}(y^{-1})\approx 70 \pm 10 (km/s.Mpc)$.

\subsection{The vacuum model with two Einstein spaces}

Here we consider the vacuum case when $T^M_P = 0$. Let us suppose
that $t_0$ is an extremum point of the function $b(t)$, i.e.
$\dot{b}(t_0) = 0.$ In this point we get $\dot{G}(t_0) = 0$. From
(\ref{6}), (\ref{7}), (\ref{8}) we get that for $t =t_0$ \bear
\label{2.6} \frac{3\ddot{a}}{a} + \frac{N\ddot{b}}{b} = 0, \mm
\label{2.7} \frac{\ddot{a}}{a} = - \frac{2 \dot{a}^2}{a^2}
-\frac{2k}{a^2}, \mm \label{2.8} \frac{\ddot{b}}{b} = -
\frac{\lambda}{b^2}. \ear

Let us suppose that we "live" near the point $t_0$, then according
to modern observations on acceleration of expansion of the
Universe (\cite{Riess,Perl}) we should put $\dot{a}(t_0) > 0$ and
 $\ddot{a}(t_0) > 0$.
This implies $k < 0$ due to (\ref{2.7}) and  $\ddot{b}(t_0) < 0,
\quad \lambda > 0$ due to (\ref{2.6}) and (\ref{2.8}). Thus, our
3-dimensional space should have negative curvature and the
internal $N$-dimensional space should have a positive curvature.

From (\ref{2.6})-(\ref{2.8}) we obtain using the definitions of cosmological
parameters
\bear
\frac{|2k|}{H_0^2 a_0^2} = 2 + |q_0|, \mm
\label{2.14}
\frac{d_2| \lambda|}{H_0^2 b_0^2} = 3 |q_0|.
\ear
Here $a_0 = a(t_0)$ and $b_0 = b(t_0)$.

Since we suppose that we "live" now near the point $t_0$, then we
get \beq \label{2.15} \frac{\dot{b}}{b} \approx \frac{\ddot{b}_0
(t - t_0)}{b_0} \eeq and due to (\ref{G-dot}) and (\ref{2.6}) we
find \beq \label{2.16} \frac{\dot{G}}{G} = - N \dot{b}/b \approx-
N \frac{\ddot{b}_0}{b_0} (t - t_0)= 3 \frac{\ddot{a}_0 (t -
t_0)}{a_0}. \eeq

The subscript "0" refers to $t_0$. Using (\ref{cosm}) we obtain in
our approximation \beq \label{2.17} \frac{\dot{G}}{G} \approx - 3
q_0 H_0^2 (t - t_0). \eeq Remind that $q_0 < 0$ and hence
$\dot{G}/G > 0$ for $t > t_0$ and $\dot{G}/G < 0$ for $t < t_0$;
 in our approximation $\dot{G}/G$ does not depend upon the
dimension of internal space $N = {\rm dim} K$.

\subsubsection{Exact $1 + 3 + 6$ solution}

Now we consider the exact solution from Ref. \cite{GIM} defined on
the manifold \beq \label{2.18} M = \R_{*} \times M^{(3)} \times
M^{(6)}, \eeq with the metric \beq \label{2.19} ds^2 = \left(f_1
f_2 \right)^{- \frac{1}{2}}[ - 2 f_1^{-2} (d\tau)^2+ |\lambda_3|
g^{(3)}_{ij}(x)dx^idx^j+ f_2 |\lambda_6| g^{(N)}_{mn}(y)dy^mdy^n]
\eeq where $(M^{(3)}, g^{(3)})$ and $(M^{(6)}, g^{(6)})$ are
Einstein spaces: \beq \label{2.20} {\rm Ric}[g^{(i)}] = \lambda_i
g^{(i)}, \eeq $i = 3, 6$. Here we use the notations: $\lambda_3 =
2k$ and $\lambda_6 = \lambda$.

In (\ref{2.19})
\bear \label{2.21}
f_1 = \left |\tau^2 + \varepsilon_3 \right|, \mm
\label{2.22}
f_2 = - 3 \varepsilon_6 \left(\tau^2 + \varepsilon_3 \right)\left[ 1 +
\tau \left( h(\tau, \varepsilon_3)+ C_1 \right) \right]+ \varepsilon_3
\varepsilon_6 > 0.
\ear
$C_1$ is constant and $\varepsilon_i = {\rm sgn}(\lambda_i)$, $i = 3, 6$, and
\bear \label{2.23}
h(\tau, \varepsilon_3)= \frac{1}{2} \ln\left|\frac{\tau-1}{\tau+1}\right|, \quad
\varepsilon_3 = - 1, \mm
\label{2.24}
h(\tau, \varepsilon_3)= \arctan(\tau), \quad \varepsilon_3 = 1.
\ear

As was mentioned above we should restrict our consideration to the
case when our 3-dimen\-si\-o\-nal space has negative curvature and
6-dimensional internal space has positive curvature, i.e. \beq
\label{2.25} \varepsilon_3 = - 1, \quad \varepsilon_6 = 1. \eeq

The analysis carried out in \cite{GIM} tells us that the scale
factor of our 3-space \beq \label{2.26} a_3 = a = (f_1
f_2)^{-1/4}|\lambda_3|^{1/2} \eeq has the minimum at some point
$\tau_{*}$ when the branch of solution with $\tau \in (\tau_{-},
\tau_{+})$ is considered. Here $\tau_{-}, \tau_{+}$ are roots of
the equation $f_2(\tau) = 0$ belonging to the interval $(0,1)$. In
this case the scale factor of our space $a_3(\tau)$ is
monotonically decreasing in the interval $(\tau_{-}, \tau_{*})$
and monotonically increasing in the interval $(\tau_{*},
\tau_{+})$.

The scale factor of internal 6-space \beq \label{2.27} a_6 = b =
(f_1 f_2)^{-1/4} f_2^{1/2} |\lambda_6|^{1/2} \eeq has maximum at
some point $\tau_{0}$. It is monotonically increasing in the
interval $(\tau_{-}, \tau_{0})$ and monotonically decreasing in
the interval $(\tau_{0}, \tau_{+})$. For other branches of
solution with either $\tau \in (\tau_{-}, \tau_{+ \infty})$ or
$\tau \in (- \infty, \tau_{+})$, $(|\tau_{-}|, |\tau_{+}| > 1)$ we
get monotonic behavior of scale factors $a_3(\tau)$ and
$a_6(\tau)$.

Let us consider the behavior of our solution in the synchronous
time \beq \label{2.28} ds^2 = - dt^2 + a_3^2(t)
g^{(3)}_{ij}(x)dx^idx^j+ a_6^2(t) g^{(N)}_{mn}(y)dy^mdy^n, \eeq
where \beq t_s= \sqrt{2} \int
^{\tau}_{\tau_{-}}d\tau^{\prime}(f_1f_2)^{-1/4} f_1^{-1} . \eeqn

The function $t_s(\tau)$ is monotonically increasing from $t_s(\tau_{-}) = 0$
to $T = t_s(\tau_{+})$.

The scale factor of 3-space has minimum at $t_{0} = t(\tau_{*})$.
$a_3(t)$ is monotonically decreasing from infinity to finite value
in the interval $(0, t_{0})$ and monotonically increasing to
infinity in the interval $(t_{0}, T)$.

The scale factor of 6-space has maximum at $t_{*} = t(\tau_{*})$.
The function $a_6(t)$ is monotonically increasing from infinity to
finite value in the interval $(0, t_{*})$ and monotonically
decreasing to infinity in the interval $(t_{*}, T)$. Only in the
case when $C_1 > 0$ we get that $t_{*} < t_0$ and hence in the
time epoch near $t_0$ we get an accelerating expansion of our
3-space.

\subsection{The model with two Ricci-flat spaces and two-component fluid}

Here we consider another example when two factor spaces are Ricci-flat.

In this case, excluding $b$ from (\ref{6}) and (\ref{8}), we get
\beq \label{14}
\frac{N-1}{3N} g^2 -g +q - \sum_{\alpha = 1}^m A^{\alpha} \Omega^{\alpha} = 0
\eeq
with
\beq \label{15}
A^{\alpha} = \frac{1}{N+2} [2N+1+3(1-N)\nu_3^{\alpha} +3N \nu_N^{\alpha}],
\eeq
where
\beq \label{16}
\nu_3^{\alpha} = p_3^{\alpha}/\rho^{\alpha}, \quad
\nu_N^{\alpha} = p_N^{\alpha}/\rho^{\alpha}, \quad \rho^{\alpha} >0\ .
\eeq

When $g$ is small we get from (\ref{14})
\beq \label{17}
g \approx q - \sum_{\alpha = 1}^m A^{\alpha} \Omega^{\alpha}.
\eeq

Note that (\ref{17}) for $N=6$, $m=1$, $\nu_3^{1}=\nu_6^{1}=0$ (so that
$A^{1}=13/8$) coincides with the corresponding relation of Wu and Wang
\cite{WW} obtained for large times in case $k=-1$ (see also \cite{IM1}).

If $k=0$, then in addition to (\ref{17}), one can obtain a separate relation
between $g$ and $\Omega^{\alpha}$, namely,
 \beq \label{18}
 \frac{N-1}{6N} g^2 - g + 1 - \sum_{\alpha = 1}^m \Omega^{\alpha} = 0
 \eeq
(this follows from the Einstein equation $R^0_0-\frac{1}{2}R=
\kappa^2T^0_0$, which is a linear combination of
(\ref{6})-(\ref{8}).

\subsubsection{Two-component example: dust + $(N-1)$-brane}

Consider now two component case: $m= 2$ \cite{IMW}. Let the first
component (matter) be a dust, i.e.
 \beq \label{19}
 \nu_3^{1} = \nu_N^{1} = 0,
 \eeq
 and the second one (quintessence) be a $(N-1)$-brane, i.e. \beq \label{20} \nu_3^{2}
  = 1, \quad \nu_N^{2} = - 1. \eeq

We remind that, as it was mentioned in \cite{IMbil}, the
multidimensional cosmological model on the product manifold $\R
\times M_1 \times \dots \times M_n$ with fields of forms (for
review see \cite{IMtop}) may be described in terms of
multicomponent "perfect" fluid \cite{IM5} with the following
equations of state for $\alpha$-s component: $p_i^{\alpha} = -
\rho^{\alpha}$ if $p$-brane worldvolume contains $M_i$ and
$p_i^{\alpha} = \rho^{\alpha}$ in the opposite case. Thus, the
field of form matter leads us either to $\Lambda$-term, or to
stiff matter equations of state in internal spaces.

In this case we get from (\ref{17}) for small $g$
\beq \label{21}
 g \approx q - \frac{2N+ 1}{N+2} \Omega^{1} + 4 \frac{N - 1}{N+2} \Omega^{2},
\eeq
 and for $k =0$ and small $g$ we obtain from (\ref{18})
\beq
 1 -g \approx \Omega^{1} + \Omega^{2}.
\eeqn

Now we illustrate these formulas by the example when $N =6$ ($K^6$
may be a Calabi-Yau manifold) and \beq \label{22}
 -q = \Omega^{1} = \Omega^{2} = 0.5.
\eeq
We get from (\ref{21})
\beq \label{23}
g \approx - \frac{1}{16} \approx -0.06
\eeq
in agreement with (\ref{18a}).

In this case the second fluid component corresponds to magnetic
(Euclidean) $NS5$-brane (in $D=10$ type I, Het or II A string
models). Here we considered for simplicity the case of the
constant dilaton field.

This example tells us that for small enough temporal variation of
$G$ we may find the estimates on G-dot without consideration of
exact solutions. But here we should select the solutions that give
us the accelerated expansion of our world. We may use, for
instance, the mechanism suggested above: but instead of curvatures
in 2 factor-spaces we should consider two fluid components.

\section{Multidimensional cosmology with anisotropic fluid \cite{AIKM,IKMN}}

\subsection{The model}

Now we consider a cosmological model describing the dynamics of
$n$ Ricci-flat spaces in the presence of $1$-component
"perfect-fluid" matter \cite{IM5}. Metric of the model \beq
 g= - \exp[2{\gamma}(t)]dt \otimes dt +\sum_{i=0}^{n} \exp[2{x^{i}}(t)] g^{i},
\eeq
is defined on the manifold
\beq \label{2.2}
 M = R \times M_{0} \times \ldots \times M_{n},
\eeq
where manifold $M_{i}$ with a metric $g^{i}$ is a Ricci-flat space of dimension
$d_i$, $i = 0, \ldots ,n $; $n \geq 2$. The multidimensional Hilbert-Einstein
equations have the following form:
\beq \label{2.3}
 R^{M}_{N}-\frac{1}{2}\delta^{M}_{N}R = \kappa^{2}T^{M}_{N},
\eeq
where $\kappa^{2}$ is the gravitational constant, and the energy-momentum
tensor is adopted as
\beq
 (T^{M}_{N})= {\rm diag}(- \rho,p_{1} \delta^{m_{1}}_{k_{1}},\ldots ,
 p_{n} \delta^{m_{n}}_{k_{n}}).
\eeq
describing anisotropic fluid in general.

We put pressures of the anisotropic perfect fluid in all spaces to
be proportional to the density \beq \label{2.8n}
 p_i(t) = (1- \frac{u_i}{d_i}) \rho(t),
\eeq
where $u_i = const$, $i = 0, \ldots ,n$. Here we put $\rho >0$.

We impose also the following restriction on vector $u = (u_i) \in
 R^{n}$
\beq \label{2.9}
 <u,u>_{*} < 0.
\eeq Here bilinear form $<.,.>_{*}: R^{n} \times R^{n} \rightarrow
R$ is defined by the relation \beq <u,v>_{*} = G^{ij} u_i v_j,
\eeq $u,v \in R^{n+1}$, where \beq \label{2.11} G^{ij} =
\frac{\delta^{ij}}{d_i} + \frac{1}{2-D} \eeq are components of
matrix inverse to a matrix of the minisuperspace metric
\cite{IM2,IMZ}
 \beq G_{ij} = d_{i} \delta_{ij} - d_{i} d_{j}.
 \eeq

In (\ref{2.11}) $D = 1 + \sum_{i=0}^{n} d_i$ is the total
dimension of  $M$ (\ref{2.2}). The restriction (\ref{2.9}) reads
\beq \label{2.13}
 <u,u>_{*} \equiv G^{ij}u_{i}u_{j}=\sum_{i= 0}^{n} \frac{(u_{i})^{2}}{d_{i}}
 + \frac{1}{2-D}(\sum_{i= 0}^{n} u_{i})^{2} < 0.
\eeq

\subsection{Solutions with power-law and exponential dependence of scale
factors}

Let us consider two special families of solutions \cite{IM5,IM10}
with the metric written in the synchronous time parametrization
\beq \label{3.1}
 g=- dt_s \otimes dt_s + \sum_{i= 0}^{n} a_i^2(t_s) g^{i}.
\eeq

These solutions have either power-law or exponential dependence of scale
factors (w.r.t. $t_s$).

\subsubsection{Solutions with power-law behavior}

The solutions with power-law behavior of scale factors take place
for \beq \label{3.2}
 <u^{(\Lambda)} - u, u>_{*} \neq 0.
\eeq

The vector $u^{(\Lambda)}_i = 2d_i$ corresponds to the
$\Lambda$-term fluid with $p_i = - \rho$ (vacuum-like). In this
case the solutions are defined by the metric (\ref{3.1}) with the
scale factors
 $a_i = {a_i}(t_s) = A_i t_s^{\nu^i},$
and the density \beq \label{3.4a}
 \kappa^2 \rho =
 \frac{- 2 < u, u>_{*}}{<u^{(\Lambda)} - u , u>_{*}^2 t_s^2},
\eeq
 \beq \label{3.5}
 \nu^i = 2 \frac{u^i}{<u^{(\Lambda)} - u , u>_{*}}
\eeq with $u^i = G^{ij} u_i$ and $A_i$ are positive constants, $i
= 0, \dots, n$.

We will use the following explicit formulas for
 contravariant components
 \beq \label{4.2}
 u^i = G^{ij} u_j = \frac{u_i}{d_i} + \frac{1}{2-D}\sum_{j =0}^{n} u_j
 \eeq
and the scalar product

 \beq \label{4.3}
 <u^{(\Lambda)} - u, u>_{*} =- \sum_{i= 0}^{n} \frac{(u_{i})^{2}}{d_{i}}
 + \frac{2}{D-2}\sum_{i = 0}^{n} u_{i}+ \frac{1}{D-2}(\sum_{i= 0}^{n} u_{i})^{2}.
 \eeq

\subsubsection{Solutions with exponential behavior}

 The solutions with exponential behavior of scale factors take
place for
 \beq \label{3.2n}
 <u^{(\Lambda)} - u, u>_{*} = 0.
 \eeq

In this case, the solutions are determined by the metric
(\ref{3.1}) with the scale factors
  $a_i = {a_i}(t_s) = A_i \exp(\nu^i t_s),$
 and the density
 $\rho = const,$
 \beq \label{3.5n}
  \nu^i = \varepsilon
      u^i\sqrt{- \frac{2 \kappa^2 \rho}{\aver{u,u}_{*}}},
  \eeq
 where
 $\varepsilon = \pm 1$, $u^i = G^{ij} u_i$ and $A_i$ are positive
constants, $i = 0, \dots, n$. Here $\rho > 0$ for
 $\aver{u,u}_{*} < 0 $ and $\rho < 0$ for $\aver{u,u}_{*} > 0 $.

{\bf Remark.} The model under consideration was integrated in
\cite{IM5} for $<u,u>_{*} < 0$. The solutions from \cite{IM5} were
generalized in \cite{IM10} to the case when a massless minimally
coupled scalar field was added. Families of exceptional solutions
with power-law and exponential behavior of scale factors in terms
of synchronous time variable were singled out in \cite{IM10} and
correspond to a constant value of the scalar field: $\varphi =
const$. When the scalar field is omitted we are lead to solutions
presented above (in \cite{IM5} these solutions were originally
written in the harmonic time parametrization). It may be verified
that the exceptional solutions with power-law dependence of scale
factors are also valid when the restriction (\ref{2.9}) is
omitted. Moreover, it may be shown that for $<u,u>_{*} = 0$ the
power-law solutions are coinciding with our vacuum Kasner-like
solution. In this case the matter source vanishes since: $\rho =
0$ in (\ref{3.4a}).

\subsection{Acceleration and variation of G}

The subspace $(M_0,g^0)$, $g^0$ to be flat, $d_0 =3$, describes
our $3$-dimensional space and $(M_i,g^i)$ internal factor-spaces.
We are interested in solutions with accelerated expansion of our
space and small enough variations of $G$ obeying experimental
tests at the present moment \cite{IMW}
 \beq \label{4.1v}
 \left|\frac{\dot{G}}{GH}\right|(t_{s0}) < 0.1,
 \eeq
 Here we
suppose that internal spaces are compact. Hence our 4-dimensions
constant is (see \cite{BIM1})
 \beq \label{4.1g} G = {\rm const}
 \prod_{i=1}^{n}( a_{i}^{-d_i}).
 \eeq

\subsubsection{Power-law expansion with acceleration}

For solutions with power-law expansion the accelerated expansion of our space
takes place for
\beq \label{4.4}
\nu^0 > 1.
\eeq

For $D=4$ when internal spaces are absent we get
 \beq
  \nu^0 = \frac{2}{6 - u_0},
 \eeq
 \beq \label{4.2aa}
  < u^{(\Lambda)} -u, u >_{*} = \frac{1}{6}(u_0 -6) u_0
 \neq 0 \eeq
that implies $u_0 \neq 0$ and $u_0 \neq 6$ (here
 $<u,u>_{*} = - \frac{1}{6} u_0^2 < 0$).
The condition $\nu^0 >1$ is equivalent to $4 < u_0 < 6$ or,
equivalently,
 \beq
  - \rho < p < - \frac{\rho}{3},
 \eeq
 that agrees with the well-known result for $D =4$.

For the power law solutions we get
\beq \label{4.5a}
 \frac{\dot{G}}{G} = - \frac{\sum_{j =1}^{n} \nu^i d_i}{t_s}, \quad H =
 \frac{\dot{a_0}}{a_0} = \frac{\nu^0}{t_s}.
\eeq
and hence
\beq \label{4.5}
 \frac{\dot{G}}{GH} = - \frac{\sum_{j =1}^{n} \nu^i d_i}{\nu^0} \equiv \delta.
\eeq

The constant parameter $\delta$ describes variations of the
gravitational constant and according to (\ref{4.1v})
 $|\delta| < 0.1.$

It follows from the definition of $\nu^i$ in (\ref{3.5}) that
\beq \label{4.7}
 \delta = - \frac{\sum_{i =1}^{n} u^i d_i}{u^0},
\eeq
or, in terms of covariant components (see (\ref{4.2}))
\beq \label{4.8}
 \delta = - \frac{(D-4) u_0 -2 \sum_{i =1}^{n} u_i}{\frac{1}{3} (5 - D)u_0 +
 \sum_{i =1}^{n} u_i}.
\eeq

Thus, the relations (\ref{4.3}), (\ref{4.4}), (\ref{4.8}) and the
constraint (\ref{3.2}) define a set of parameters $u_i$ compatible
with the acceleration and tests on G-dot.

In what follows we will show that these relations do really define a non-empty
set of parameters $u_i$ describing equations of state.

\paragraph{The case of constant $G$.}

Consider now the most important case $\delta =0$, i.e. when the
variation of $G$ is absent: $\dot{G} = 0$. Indeed, there is a
tendency of lowering the upper bound on $G$, e.g. according to
arguments of \cite{BZhuk} $\delta < 10^{-4}$. This constraint just
follows from the identity \beq \label{4.Al} \frac{\dot{G}}{G}
=\frac{\dot{\alpha}}{\alpha} \eeq that may take place in some
multidimensional models ($\alpha$ is the fine structure constant).

\subparagraph{Isotropic case.}

First, we consider an isotropic case when pressures in all
internal spaces are coinciding. This takes place when \beq
\label{4.9}
 u_i = v d_i,
\eeq
$i = 1, \ldots, n$. For pressures in internal spaces we get from (\ref{2.8n})
\beq \label{4.10}
 p_i = (1- v) \rho,
\eeq
$i = 1, \ldots, n$.
In the isotropic case we get from (\ref{2.13}) and (\ref{4.3})
\beq \label{4.11a}
 <u, u>_{*} = \frac{1}{2-D} [- \frac{1}{3} (d-1) u_0 + 2d u_0 v- 2d
 v^2],
\eeq

\beq \label{4.11b} <u^{(\Lambda)} - u, u >_{*} = \frac{1}{2-D}[2
u_0 + 2dv + \frac{1}{3}(d-1) u_0^2- 2d u_0 v+ 2d v^2 ].
 \eeq

For $\delta = 0$ we get in the isotropic case $v = \frac{u_0}{2}$
or in terms of pressures

\beq \label{4.12p} p_i= \frac{1}{2}( 3 p_0 - \rho), i = 1, \ldots,
n. \eeq

Substituting into (\ref{4.11a}) and (\ref{4.11b}) we get \beq
\label{4.13a} <u, u>_{*} = - \frac{1}{6} u_0^2, \eeq

\beq \label{4.13b}
<u^{(\Lambda)} - u, u >_{*} = \frac{1}{6} u_0 (u_0 -6).
\eeq

Remarkably, we obtain the same relations as in $D = 4$ case (see Remark above).
For our solution we should put $u_0 \neq 0$ and $u_0 \neq 6$.

Using (\ref{4.9}) we get $u^0 = - u_0/6$ and $u^i = 0$ for $i
 > 0$, hence $ \nu_i =0$ for $i = 1, \ldots, n$, i.e. all internal spaces are
static.

Metric (\ref{3.1}) reads in this case \beq \label{4.14}
 g=- dt_s \otimes dt_s + A_0^2 t_s^{2 \nu^0} g^{0} + \sum_{i= 1}^{n} A_i^2 g^{i},
\eeq
where $A_i$ are positive constants, and
\beq \label{4.15}
 \nu^0 = \frac{2}{6 - u_0}.
\eeq

We see, that the power $\nu^0$ is the same as in $D =4$ case. For the density
we get from (\ref{3.4a})
\beq \label{4.16}
 \kappa^2 \rho = \frac{12}{(u_0 -6)^2 t_s^2}.
\eeq

Thus, equations of state (\ref{2.8n}) with relations (\ref{4.9})
imposed, lead to the solution (\ref{4.14})-(\ref{4.16}) with
Ricci-flat (e.g. flat) 3-metric and $n$ static internal Ricci-flat
spaces. For $4 < u_0 < 6,$
 or $- \rho < p_0 < - \frac{\rho}{3}$, we get an
accelerated expansion of our 3-dimensional Ricci-flat space.

\subparagraph{Non-isotropic case.}

Let us consider the anisotropic (w.r.t. internal spaces) case with
$\delta = 0$, or, equivalently (see (\ref{4.8})) , \beq
\label{4.18a} (D-4) u_0 = 2 \sum_{i =1}^{n} u_i. \eeq

This implies
 \beq \label{4.18}
  <u^{(\Lambda)} - u, u >_{*} = \frac{1}{6} u_0 (u_0 -6) - \Delta,
 \eeq
 \beq
   <u, u >_{*} = - \frac{1}{6} u_0^2 + \Delta,
 \eeq
where
 \beq \label{4.19}
  \Delta= \sum_{i=1}^n \frac{u_i^2}{d_i} - \frac{1}{d}\left(\sum_{i=1}^n
  u_i\right)^2 \geq 0.
 \eeq

The inequality in (\ref{4.19}) could be readily proved using the
well-known Cauchy-Schwartz inequality:
 \beq \label{4.19CS}
 (\sum_{i=1}^n b_i^2) (\sum_{i=1}^n c_i^2)\geq (\sum_{i=1}^n b_i c_i )^2.
 \eeq

Indeed, substituting $b_i = \sqrt{d_i}$ and $c_i = u_i/ \sqrt{d_i}$ into
 (\ref{4.19CS}) we get (\ref{4.19}). The equality in (\ref{4.19CS}) takes place
only when the vectors $(b_i)$ and $(c_i)$ are linearly dependent that for our
choice reads: $u_i/ \sqrt{d_i} = v \sqrt{d_i}$ where $v$ is constant. Thus,
 $\Delta = 0$ only in the isotropic case (\ref{4.9}). In the non-isotropic case
we get $\Delta > 0$.

In what follows we will use the relation
 \beq \label{4.19a}
 <u^{(\Lambda)} - u, u >_{*} =\frac{1}{6} (u_0 - u_0^{+})(u_0 - u_0^{-}),
 \eeq
where
 \beq \label{4.19b}
 u_0^{\pm} = 3 \pm \sqrt{9 + 6 \Delta}
 \eeq
are roots of quadratic polinomial (\ref{4.18}), obeying $u_0^{-} <
0, \quad u_0^{+} > 6$ for $\Delta > 0$. It follows from
(\ref{4.18a}) that $u^0 = - u_0/6$ and hence
 \beq \label{4.20}
 \nu^0 = - \frac{2u_0}{u_0 (u_0 -6) - 6\Delta}, \quad u_0 \neq u_0^{\pm}.
 \eeq

The function $\nu^0 (u_0)$ is monotonically increasing: i) from $ 0$ to $+
 \infty$ in the interval $(- \infty, u_0^{-})$; ii) from $-\infty$ to $+ \infty$
in the interval $(u_0^{-}, u_0^{+})$; iii) from $-\infty$ to $0$
in the interval $(u_0^{+}, + \infty)$.

The accelerated expansion of our space takes place when $\nu^0 >
1$, or, equivalently, when
 \beq \label{4.22a}
 ( A) \quad u_0 \in (u_{0*}^{-}, u_0^{-}),
 \eeq
 \beq \label{4.22b}
 (B) \quad u_0 \in (u_{0*}^{+}, u_0^{+}).
 \eeq
 \beq \label{4.23}
 u_{0*}^{\pm} = 2 \pm \sqrt{4 + 6 \Delta}.
 \eeq

In terms of $w_0$-parameter:
 \beq \label{4.24}
  p_0 = w_0 \rho, \quad w_0 = 1 - \frac{u_0}{3},
 \eeq
these two branches read:

 \beq \label{4.25a}
 (A) \quad w_0^{-} = \sqrt{1 + \frac{2}{3} \Delta} < w_0< \frac{1}{3} +
 \frac{2}{3} \sqrt{1 + \frac{3}{2} \Delta}= w_{0*}^{-}
 \eeq

 \beq \label{4.25b}
 (B) \quad w_0^{+} = - \sqrt{1 + \frac{2}{3} \Delta} < w_0< \frac{1}{3} -
 \frac{2}{3} \sqrt{1 + \frac{3}{2} \Delta} = w_{0*}^{+}.
 \eeq

The first branch $(A)$ describes a superstiff matter ($w_0 > 1$)
with negative density. Indeed, the relation $\rho < 0$ follows
from (\ref{3.4a}) and $<u, u  >_{*} > 0$, see (\ref{4.18a}).

The second branch $(B)$ corresponds to matter with a broken weak
energy condition (since $w_0 < - \frac{1}{3}$) and positive
density (since $<u, u >_{*} < 0$). This matter is a fantom one
(i.e. $w_0 < - 1$) when $\Delta \geq 2$. For $\Delta < 2$ the
interval $(w_0^{+}, w_{0*}^{+})$ contains both fantom points
 ($w_0 < - 1$) and non-fantom ones ($w_0 > - 1$).

\paragraph{The case of varying $G$.}

Now we consider another important case $\delta \neq 0$, i.e. when
the variation of $G$ is non-zero: $\dot{G} \neq 0$. We use the
bound $|\delta| < 0.1$, stating the smallness of $\delta$. Using
(\ref{4.8}) we get
 \beq \label{5.18a}
 \sum_{i =1}^{n} u_i = \frac{1}{2} d b u_0 ,
 \eeq
where $d = D-4$ and
 \beq \label{5.18bd}
 b = b(\delta) = \frac{1+ \frac{1-d}{3d} \delta}{1 -\frac{\delta}{2}}.
 \eeq

For the scalar product we get from (\ref{5.18a})
\beq \label{5.18}
 <u^{(\Lambda)} - u, u >_{*} =\frac{A}{6} u_0^2 - B u_0 - \Delta,
\eeq
 \beq \label{5.18b}
 <u, u >_{*} = - \frac{A}{6} u_0^2 + \Delta,
\eeq where $\Delta$ was defined in (\ref{4.19}) (see (\ref{2.13})
and (\ref{4.3})), and

 \beq \label{5.18cc}
 A = A(\delta) = 1 - \frac{(d+2) \delta^2}{12d (1 -\frac{\delta}{2})^2},
\eeq
 \beq \label{5.18dd}
 B = B(\delta) = \frac{1 - \frac{\delta}{3}}{1 - \frac{\delta}{2}}.
\eeq
It should be noted that due to $|\delta| < 0.1$ $A$ is positive $A > 0$ and
close to $1$: $|A -1| < \frac{1}{3} 10^{-2}$.

For contravariant component $u^0$ we get from (\ref{4.2}) and (\ref{5.18a})
\beq \label{5.18f}
 u^0 = - \frac{C}{6} u_0,
\eeq
\beq \label{5.18ff}
 C = C(\delta) = 3B - 2 = \frac{1}{1 - \frac{\delta}{2}}.
\eeq

It follows from (\ref{5.18}) and (\ref{5.18f}) that
(see (\ref{3.5}))

\beq \label{5.20}
 \nu^0 = - \frac{2 C u_0}{\frac{A}{6} u_0^2 - B u_0 - \Delta}.
\eeq Here $u_0 \neq u_0^{\pm}$, where

\beq \label{5.19b}
 u_0^{\pm} = u_0^{\pm}(\delta) = \frac{1}{A}(3 B \pm \sqrt{9 B^2 + 6 A \Delta} )
\eeq
are roots of quadratic polinomial (\ref{5.18}).

In what follows we will use the identity
\beq \label{5.20c}
 \nu^0 -1 = - \frac{A u_0^2 - 4 u_0 - 6 \Delta}{A u_0^2 - 6 B u_0 - 6 \Delta}.
\eeq

 \subparagraph{Isotropic case.}

Let us consider an isotropic case (\ref{4.9}). In this case we obtain from
(\ref{5.18a})
\beq \label{5.12}
 v = \frac{1}{2} d b u_0 .
\eeq
or, in terms of pressures
\beq \label{5.12p}
 p_i= \frac{1}{2} [ 3b p_0 + (2 -3b) \rho ], i = 1, \ldots, n.
\eeq
 For scalar products we get
 \beq \label{5.13a}
 <u, u>_{*} = - \frac{A}{6} u_0^2,
\eeq
 \beq \label{5.13b}
 <u^{(\Lambda)} - u, u >_{*} = \frac{A}{6} u_0 (u_0 - 6 B).
\eeq

For our solution we should put $u_0 \neq 0$ and $u_0\neq 6B/A$.

The metric (\ref{3.1}) reads in our case
\beq \label{5.14}
g= - dt_s \otimes dt_s +A_0^2 t_s^{2 \nu^0} g^{0} +t_s^{2 \nu}
\sum_{i= 1}^{n} A_i^2 g^{i},
\eeq
where $A_i$ are positive constants,
 \beq \label{5.15}
 \nu^0 = - \frac{2C}{A u_0 - 6B},
 \eeq
\beq \label{5.15a}
 \nu = \nu^i = \frac{2 \delta}{d(1 - \frac{\delta}{2}) (A u_0 - 6B)},
\eeq
 $i = 1, \ldots, n$. The last formula follows from (\ref{3.5}) and
 \beq \label{5.15b}
 u^i = \frac{u_0}{6d(1 - \frac{\delta}{2})}.
 \eeq

We see, that the power $\nu^0$ is not coinciding for $\delta \neq 0$ with that
from $D =4$ case.

For the density, since $A > 0$, we get from (\ref{3.4a})
 \beq \label{5.16}
 \kappa^2 \rho = \frac{12A}{(u_0 -6B)^2 t_s^2} > 0.
 \eeq

The condition of accelerated expansion of 3-dimensional space
$\nu^0 > 1$ reads as
 \beq \label{5.17}
 \frac{4}{A(\delta)} < u_0 < \frac{6B(\delta)}{A(\delta)}
 \eeq
or, equivalently, in terms of $w_0$-parameter ($p_0 = w_0 \rho$) (\ref{4.24})
 \beq \label{5.17a}
  w_{0}^{+}(\delta) = 1 - \frac{2B(\delta)}{A(\delta)} < w_0 < 1 -
  \frac{4}{3A(\delta)} = w_{0*}^{+}(\delta).
 \eeq

For $\delta > 0$ we get an isotropic contraction of total internal space $M_{1}
\times \ldots \times M_{n}$. In this case $w_{0}^{+}(\delta) < - 1$ and hence
phantom matter may occur with the equation of state close to the vacuum one,
since
\beq \label{5.17b}
 w_{0}^{+}(\delta) + 1 = - \frac{\delta (1 +\frac{\delta}{d})}{3(1 -
 \delta + \frac{(d -1) \delta^2}{6d})}.
 \eeq
For small $\delta$ we have $w_{0}^{+}(\delta) = 1 - \frac{\delta}{3} +
 O(\delta^2)$.

For $\delta < 0$ we get an isotropic expansion of total internal space. In this
case $w_{0}^{+}(\delta) > - 1$ and the phantom matter does not occur. In both
cases $w_{0*}^{+}(\delta) < - \frac{1}{3}$ and $w_{0*}^{+}(\delta) +
 \frac{1}{3} = O(\delta^2)$.

\subparagraph{Non-isotropic case.}

Now we consider a non isotropic case $\Delta > 0$ when $\delta \neq 0$.

Using relation (\ref{5.20c}) we obtain
 \beq \label{5.20a}
 \nu^0 -1 = - \frac{(u_0 - u_{0*}^{+})(u_0 - u_{0*}^{-})}{(u_0 - u_0^{+})
 (u_0 - u_0^{-})}.
 \eeq
where $u_0^{\pm} = u_0^{\pm}(\delta)$ were defined in (\ref{4.19}) and
 \beq \label{5.19c}
 u_{0*}^{\pm} = u_{0*}^{\pm}(\delta) =2 \pm \sqrt{4 + 6 A \Delta}.
 \eeq

The accelerated expansion of 3-dimensional space takes place when
$\nu^0 > 1$, i.e.

 $(A) \quad u_0 \in (u_{0*}^{-}(\delta), u_0^{-}(\delta))$,
$\quad (B) \quad u_0 \in (u_{0*}^{+}(\delta), u_0^{+}(\delta))$.

 In terms of $w_0$-parameter $p_0 = w_0 \rho$, ($w_0 = 1 -
\frac{u_0}{3}$) these two branches read:
 \beq \label{5.25a}
 (A) \quad w_0^{-}(\delta) < w_0 < w_{0*}^{-}(\delta), \quad
 (B) \quad w_0^{+}(\delta) < w_0 < w_{0*}^{+}(\delta),
 \eeq
where
\beq \label{5.26a}
  w_0^{\pm}(\delta) = 1 - \frac{u_0^{\pm}(\delta)}{3}, \quad
w_{0*}^{\pm}(\delta) = 1 - \frac{u_{0*}^{\pm}(\delta)}{3}.
 \eeq

For small $\delta$ we have
 \beq \label{5.27a}
 w_0^{\pm}(\delta) = w_0^{\pm}(0) -\frac{\delta}{6} (1 \pm \frac{3}{\sqrt{9 +
 6 \Delta}}) + O(\delta^2),
 \eeq

 \beq \label{5.27b}
 w_{0*}^{\pm}(\delta) = w_{0*}^{\pm}(0) + O(\delta^2).
 \eeq

Thus, for small $\delta$ the lower and upper bounds on $w_0$ have a small
deviation from those obtained in the case $\delta = 0$. For small $\delta$ the
upper bounds shift only on $O(\delta^2)$ term while the lower bounds shift on
 $O(\delta)$ term.

The first branch $(A)$ describes a superstiff matter $w_0 > 1$, since
 $w_0^{-}(\delta) > 1$ due to $u_0^{-}(\delta) < 0$. It may be shown that the
density is negative in this case since $<u, u >_{*} > 0$.

For the second branch $(B)$ we get for the upper bound
$w_{0*}^{+}(\delta) < - 1/3$ due to $u_{0*}^{+}(\delta) > 4$. For
the lower bound we find that $w_{0}^{+}(\delta) < - 1$ only if
  \beq \label{5.28c}
  \Delta > 6(A(\delta) -B(\delta)) =- \frac{\delta}{(1
 - \frac{\delta}{2})^2}.
  \eeq

This is the condition on appearance of the phantom matter. For
$\delta
> 0$ this inequality is valid, but for $\delta < 0$ it is
satisfied only for big enough $\Delta$.


\subsubsection{Exponential expansion with acceleration}

For solutions from 4.2.2, an accelerated expansion of our space
takes place for $\nu^0 > 0$. For $D=4$, when internal spaces are
absent, we get $u^0 = - u_0/6$ and $\aver { u, u }_{*} = - \frac
{1}{6} u_0^2$,
  $\aver {u^{(\Lambda)} -u, u }_{*} = \frac   {1}{6}(u_0 -6) u_0 = 0$,
which implies $u_0 = 6$, or, equivalently, $p = -  \rho$.

 We get \beq \label{4.4a} \nu^0 = - \varepsilon \sqrt{
 \frac{\kappa^2 \rho}{3}}, \eeq
which agrees with the well-known
result for $D =4$ de-Sitter solution with cosmological constant
 $\Lambda = \kappa^2 \rho > 0$. The condition $\nu^0 >0$ is
equivalent to $\varepsilon = -1$.

For our exponential solutions we get
 \beq
 \frac{\dot{G}}{G} = - \sum_{j =1}^{n} \nu^i d_i, \quad
 H = \frac{\dot{a_0}}{a_0} = \nu^0,
 \eeq
and hence
 \beq \label{4.5n}
 \dot{G}/(GH) =- \frac{1}{\nu^0} \sum_{j =1}^{n} \nu^i d_i \equiv \delta, i.e.
  \eeq
we get the same relation as in (\ref{4.5}).

 The constant parameter $\delta$ describing variation of the
gravitational constant, obey the restriction $|\delta| < 0.1$.

Due to (\ref{4.5n}) we get the same relations (\ref{4.7}) and
(\ref{4.8}) for $\delta$ as in the power-law case.

\paragraph{The case of constant $G$. Isotropic case.}

Consider the important case $\delta =0$, i.e., when the variation
of $G$ is absent: $\dot{G} = 0$.

First, we consider the isotropic case when pressures coincide in
all internal spaces, see (\ref{4.9}). Here, we obtain the same
relations as in $D = 4$ case. For our solution, we should put $u_0
\neq 0$ and hence, due to (\ref{3.2n}),
 $ u_0 = 6$, i.e. $ p_0= - \rho$.

Using (\ref{4.9}), we get $u^0 = - u_0/6 = -1$ and $u^i = 0$ for
$i > 0$, hence $ \nu_i =0$ for $i = 1, \ldots, n$, i.e., all
internal spaces are static.

The metric (\ref{3.1}) reads in this case as
 \beq
 g=- dt_s \otimes dt_s + A_0^2 \exp (2 \nu^0 t_s) g^{0} + \sum_{i= 1}^{n} A_i^2 g^{i},
 \eeq
where $A_i$ are positive constants, and
 \beq
 \nu^0 = - \varepsilon \sqrt{\frac{\kappa^2 \rho}{3}}.
 \eeq
For accelerated expansion we get $\varepsilon = -1$. We see that
the power $\nu^0$ is the same as in $D =4$ case.

\paragraph{Anisotropic case.}

Consider  now the anisotropic (w.r.t. internal spaces) case with
 $\delta = 0$, or, equivalently when (\ref{4.18a}) is satisfied.
It follows from (\ref{4.18a}) that $u^0 = - u_0/6$ and hence \beq
 \nu^0 = - \varepsilon \frac{u_0}{6} \sqrt{\frac{12 \kappa^2 \rho}
 {u_0^2 - 6\Delta}}, \quad u_0 = u_0^{\pm}.
\eeq
The accelerated expansion of our space takes place when
$\nu^0 > 0$, or, equivalently, when either \bear
 {\bf (A)} \quad u_0 = u_0^{-}, \quad \varepsilon = 1 \quad {\rm or} \quad
 {\bf (B)} \quad u_0 = u_0^{+}, \quad \varepsilon = -1.
\ear

In terms of the parameter $w_0$, see (\ref{4.24}), these two
branches read:
 \bear
 {\bf (A)} \quad w_0 = w_0^{-} = \sqrt{1 + \fract{2}{3} \Delta}, \mm
 {\bf (B)} \quad w_0 = w_0^{+} = - \sqrt{1 + \fract{2}{3} \Delta}.
 \ear
The first branch (A) describes the super-stiff matter ($w_0 > 1$)
with negative density $\rho < 0$.

The second branch (B) corresponds to matter with positive density (since
 $\aver{u, u }_{*} < 0$). This matter is the phantom one (i.e., $w_0 < - 1$) when
 $\Delta > 0$.

\paragraph{The case of varying $G$.}

Now, we consider the case $\delta \neq 0$, i.e., when $\dot{G}
\neq 0$. We take the observational bound $|\delta| < 0.1$. Using
(\ref{4.8}), we get relations (\ref{5.18a}) and (\ref{5.18bd}). It
follows from (\ref{5.18}) and (\ref{5.18f}) that

 \beq
  \nu^0 = - \varepsilon \frac{C u_0}{6} \sqrt{\frac{12 \kappa^2 \rho}
  {A u_0^2 - 6\Delta}}.
 \eeq
 Here $u_0 = u_0^{\pm}(\delta)$ are defined in (\ref{5.19b}).

\subparagraph{Isotropic case.}

Let us consider the isotropic case (\ref{4.9}). We should put $u_0
\neq 0$ and hence $u_0 = 6B/A > 0$. The metric (\ref{3.1}) reads
 \beq
 g= - dt_s \otimes dt_s + A_0^2 \exp(2 \nu^0 t_s) g^{0} +\exp(2 \nu t_s)
 \sum_{i= 1}^{n} A_i^2 g^{i},
 \eeq
where $A_i$ are positive constants,
 \bear
 \nu^0 = - \varepsilon\frac{C u_0}{6} \sqrt{\frac{12 \kappa^2
 \rho}{A u_0^2}},\quad {\rm and} \mm
 \nu = \nu^i = \varepsilon\frac{\delta u_0}{6 d(1 - \delta/2)}
 \sqrt{\frac{12 \kappa^2 \rho}{A u_0^2}}, \quad i = 1, \ldots, n.
 \ear
We see that the power $\nu^0$ does not coincide for $\delta \neq
0$ with that in $D =4$ case.

The accelerated expansion condition for 3D space, $\nu^0 > 0$,
reads as
 \beq
 u_0 = \frac{6B(\delta)}{A(\delta)}, \quad \varepsilon = -1
 \eeq
or, equivalently, in terms of $w_0$ (\ref{4.24}) ($p_0 = w_0
 \rho$) \beq
 w_0 = w_{0}^{+}(\delta) = 1 - \frac{2B(\delta)}{A(\delta)}.
\eeq

For $\delta > 0$, we get an isotropic contraction of the whole
internal space $M_{1} \times \ldots \times M_{n}$. In this case
$w_{0}^{+}(\delta) < - 1$ and hence phantom matter occurs with the
equation of state close to the vacuum one since
 \beq
 w_{0}^{+}(\delta) + 1 = - \frac{\delta (1 + \delta/d)}{3[1 - \delta + (d -1)
 \delta^2/(6d)]}.
 \eeq
For small $\delta$ we have $w_{0}^{+}(\delta) = -1 - \delta/3 +
  O(\delta^2)$.

For $\delta < 0$ we get an isotropic expansion of the whole
internal space. Then, $w_{0}^{+}(\delta) > - 1$, and phantom
matter does not occur.

\subparagraph{Anisotropic case.}

Consider the anisotropic case $\Delta > 0$ when $\delta \neq 0$.
Here
 $u_0 = u_0^{\pm}(\delta)$, see (\ref{5.19b}).

Accelerated expansion of 3-dimensional space takes place when
$\nu^0 > 0$, or, equivalently, when either \bear
 {\bf (A)} \quad u_0 = u_0^{-}(\delta), \quad \varepsilon = 1 \quad {\rm
 or} \quad
 {\bf (B)} \quad u_0 = u_0^{+}(\delta), \quad \varepsilon = - 1 .
\ear

In terms of the parameter $w_0$ ($p_0 = w_0 \rho$, $w_0 = 1 - \frac{u_0}{3}$)
these two branches read:
\bear
 {\bf (A)} \quad w_0 = w_0^{-}(\delta), \quad
 {\bf (B)} \quad w_0 = w_0^{+}(\delta),
\ear
where
 $w_0^{\pm}(\delta) = 1 - {u_0^{\pm}(\delta)}/{3}$.

For small $\delta$ (see (\ref{5.27a}) the parameter
 $w_0^{\pm}(\delta)$ has a small deviation from that obtained for
 $\delta = 0$.

The branch (A) describes super-stiff matter $w_0 > 1$ since
 $w_0^{-}(\delta) > 1$ due to $u_0^{-}(\delta) < 0$. It may be shown that the
density is negative in this case since $\aver{u, u }_{*} > 0$.

For branch (B) we find that $ w_{0}^{+}(\delta) < - 1 $  only if
 (\ref{5.28c}) is satisfied.

This is the condition on appearance of the phantom matter. For
$\delta > 0$ this inequality is valid, but for $\delta < 0$ it is
satisfied only for a big enough value of anisotropy parameter
$\Delta$, see (\ref{5.28c}).

 \section{$S$-brane solution with acceleration and small
 variation of $G$}

 \subsection{The model}

In this section we deal with $S$-brane solutions describing two
electric branes and a set of $l$ scalar fields \cite{AIM-07}.

The model is governed by the action
\bear \label{1.1}
S=\int d^Dx \sqrt{|g|} \{R[g] - h_{\alpha\beta} g^{MN}\p_M\varphi^\alpha
\p_N\varphi^\beta \nn
-\sum_{a = 1,2}\frac{1}{N_a!}\exp[2\lambda_a(\varphi)](F^a)^2 \}.
\ear
Here $g=g_{MN}(x)dx^M\otimes dx^N$ is a metric of pseudo-Euclidean signature
$(-,+, \dots, +)$, $F^a = dA^a$ is a form of rank $N_a$, $(h_{\alpha\beta})$ is
non-degenerate symmetric matrix, $\varphi=(\varphi^\alpha) \in \R^l$ is a
vector of $l$ scalar fields, $\lambda_a(\varphi)=\lambda_{a
\alpha}\varphi^\alpha$, is a linear function. Here $a = 1,2$ and $\alpha, \beta
=1, \dots, l$. In (\ref{1.1}) $|g| = |\det (g_{MN})|$.

We consider as an example the manifold \beq \label{1.2} M = (0, +
\infty) \times M_1 \times M_2 \times M_3 \times M_4 \times M_{5}.
\eeq where $M_i $ are oriented Riemannian Ricci-flat spaces of
dimension $d_i$, $i = 1, \dots, 5$, and $d_1 = 1$.

Let two electric branes be defined by sets $I_1 = \{ 1, 2, 3 \}$ and $I_2 = \{
1, 2, 4 \}$. They intersect on $M_1 \times M_2$. The first brane covers also
$M_3$ and the second one covers $M_4$. The first brane corresponds to the form
$F^1$, and the second one corresponds to the form $F^2$.

For world-volume dimensions of branes we get $d(I_s) = N_s -1 = 1
+ d_2 + d_{2 + s}$, $s=1,2$ and $d(I_1 \cap I_2) = 1 + d_2$ is the
dimension of branes intersection.

Consider now $S$-brane solution governed by the function $\hat{H}
= 1 + P \rho^2$, where $\rho$ is a time variable, $P = K Q^2/8$,
$s = 1,2$.
\beq \label{1.7} K = K_{s} = d(I_s)\bigl( 1+
\frac{d(I_s)}{2-D}\bigr)+ \lambda_{s \alpha } \lambda_{s \beta}
h^{\alpha \beta}, \eeq
$s = 1,2$ is supposed to be non zero. Thus,
$K_1 = K_2 = K$. Here $(h^{\alpha \beta}) = (h_{\alpha
\beta})^{-1}$.

The branes intersection rule is following one \beq \label{1.10}
d(I_1 \cap I_{2}) = \frac{d(I_1)d(I_{2})}{D -2} +\lambda_{1 \alpha
} \lambda_{2 \beta } h^{\alpha\beta} - \frac{1}{2} K. \eeq

This relation corresponds to Lie algebra $A_2$ \cite{IMJ,IMtop}.
Remind that $K_s = (U^s,U^s)$, $s = 1,2$, where "electric" $U^s$
vectors and scalar products were defined in \cite{IM11,IM12} (see
also \cite{IMC,IMJ}). Relations $K_1 = K_2$ and (\ref{1.10})
follow just from the formula $(A_{s s'}) =
(2(U^s,U^{s'})/(U^{s'},U^{s'})$, where $(A_{s s'})$ is the Cartan
matrix for $A_2$ (with $A_{12} = A_{21} = -1$).

We consider the following exact solution
 \bear \label{1.11}
 g= \hat{H}^{2 A}\left\{ - d\rho \otimes d \rho+ \hat{H}^{-4 B}
 ( \rho^2 g^1 + g^2)+ \hat{H}^{-2 B} g^3 + \hat{H}^{-2 B} g^4 + g^5 \right\}, \mm
 \label{1.12}
 \exp(\varphi^\alpha)= \hat{H}^{ B \lambda_{1}^{\alpha} + B \lambda_{2}^{\alpha}}, \mm
 \label{1.13a}
 F^1= - Q \hat{H}^{-2}\rho d\rho \wedge \tau_1 \wedge \tau_2 \wedge \tau_3, \mm
 \label{1.13b}
 F^2= - Q \hat{H}^{-2}\rho d\rho \wedge \tau_1 \wedge \tau_2 \wedge \tau_4,
 \ear
where
 \bear \label{1.14A}
  A = 2 K^{-1} \sum_{s = 1,2} \frac{d(I_s)}{D-2}, \mm
  \label{1.14B}
  B = 2 K^{-1},
 \ear
$s = 1,2$. Here $g_1 = dx \otimes dx$, $\tau_1 = dx$ and $\tau_i$ denotes a
volume form on $M_i$. Remind that all Ricci-flat metrics $g^2, \ldots, g^5$
have Euclidean signatures.

This solution is a special case of a more general solution from
\cite{GIM-flux} corresponding to Lie algebra $A_2$.

\subsection{Solutions with acceleration}

Let us introduce a "synchronous" time variable $\tau = \tau(\rho)$
by the following relation: \beq \label{2.1} \tau = \int_{0}^{\rho}
d \bar{\rho} [\hat{H}(\bar{\rho})]^{A} \eeq

We put $P < 0$ and hence $K < 0$ that implies $A < 0$. Consider
two intervals of the parameter $A$: $(i) \ A < -1$ and $(ii) \ -1
< A < 0$.

For the case $(i)$ the function $\tau = \tau (\rho)$ is
monotonically increasing from $0$ to $+ \infty$, for $\rho \in (0,
\rho_1)$, where $\rho_1 = |P|^{-1/2}$, while for the case $(ii)$
it is monotonically increasing from $0$ to the finite value
$\tau_1 = \tau(\rho_1)$.

Let space $M_5$ be our 3-dimensional space with a scale factor
$a_5 = \hat{H}^A$.

For the  branch $(i)$ we get an asymptotical relation $a_5 \sim
{\rm const} \ \tau^{\nu}$,  for $\tau \to +\infty $, where \beq
\label{2.4n} \nu = \frac{A}{A+1} \eeq and $\nu > 1$. For $(ii)$ we
obtain $a_5 \sim {\rm const} \ (\tau_1 - \tau)^{\nu}$,  for $\tau
\to \tau_1 - 0$, where $\nu < 0$ due to (\ref{2.4n}). Thus, we get
an asymptotical accelerated expansion of 3-dimensional factor
space $M_5$ in both cases i) and ii) and $a_5 \to + \infty$.

Moreover, it may be readily verified that the accelerated expansion takes place
for all $\tau > 0$, i.e.

$\dot{a_5} > 0, \quad \ddot{a_5} > 0$. Here and in what follows we
denote $\dot{f} = df/d \tau$.

Indeed, using the relation $d\tau/d \rho = \hat{H}^A$ (see
(\ref{2.1})) we get \beq \label{2.6a} \dot{a_5} = \frac{d \rho}{d
\tau} \frac{d a_5}{d \rho} =\frac{2|A||P|\rho} {\hat{H}}, \eeq and
\beq \label{2.6b} \ddot{a_5} = \frac{d \rho}{d \tau} \frac{d}{d
\rho}\frac{da_5}{d \rho} = \frac{2|A||P}{\hat{H}^{2 + A}} (1 + |P|
\rho^2), \eeq that certainly implies inequalities for derivatives
of scale-factor $a_5$.

Now we consider the variation of $G$. For our model the
4-dimensional gravitational coupling (in Jordan frame) is \beq G =
\const \cdot \prod\nolimits_{i=1}^{4}( a_{i}^{-d_i}) =
\hat{H}^{2A} \rho^{-1}, \eeq where \beq \label{2.7a } a_1 =
\hat{H}^{A - 2B} \rho, \quad a_2 = \hat{H}^{A - 2B}, \quad a_3
=a_4 = \hat{H}^{A - B} \eeq are scale factors of "internal" spaces
$M_1, \dots, M_4$, respectively.

The function $G({\tau})$ has minimum at the point $\tau_0$
corresponding to \beq \rho_{0} = \frac{|P|^{-1}}{1 +4 |A|}. \eeq
At this point the variation of $G$ is zero. This follows from
explicit relation for dimensionless variation of $G$ \beq
\label{2.11n} \delta = \dot{G}/(GH) = 2 + \frac{1-|P|\rho^2}{2
A|P| \rho^2}, \eeq where $H = \frac{\dot{a_5}}{a_5}$ is the Hubble
parameter. The function $G({\tau})$ is monotonically decreasing
from $+ \infty$ to $G_0 = G(\tau_0)$ for $\tau \in (0, \tau_0)$
and monotonically increasing from $G_0 = G(\tau_0)$ to $+ \infty$
for $\tau \in (\tau_0, \bar{\tau}_1)$. Here $\bar{\tau}_1 =
+\infty $ for the case i) and $\bar{\tau}_1 = \tau_1$ for the case
ii). The scale factors $a_2(\tau), a_3(\tau), a_4(\tau)$ are
monotonically increasing from $1$ to $0$ for $\tau \in (0,
\bar{\tau}_1)$, since the powers $A-B$ and $A-2B$ are positive and
$P < 0$. The scale factor $a_1(\tau)$, is monotonically decreasing
from zero to $a_1(\tau_2)$ for $\tau \in (0, \tau_2)$ and
monotonically increasing from $a_1(\tau_2)$ to zero for $\tau \in
(\tau_2, \bar{\tau}_1)$, where $\tau_2$ is the point of maximum.

We should treat only solutions with accelerated expansion of our
space and small enough variations of the gravitational constant
obeying the present experimental constraint \beq \label{2.10}
|\delta| < 0.1. \eeq Here like in case of the model with two
curvatures \cite{DIKM} $\tau$ is restricted by the interval
containing $\tau_0$. It follows from (\ref{2.11n}) that in the
asymptotical regions $\delta \to 2$ that does not agree with
experimental bounds (\ref{2.10}). This restriction is satisfied
for the interval containing the point $\tau_0$ where $\delta = 0$.

The calculation of G-dot in the linear approximation near
$\tau_0$, gives the following approximate relation for
dimensionless parameter of reciprocal variation of $G$
 \cite{AIM-07}
   \beq \label{2.13n} \delta \approx (8 + 2 |A|^{-1})
    H_0 (\tau - \tau_0),
   \eeq
where $H_0 = H(\tau_0)$ (compare with the analogous relation in
\cite{DIKM}). This relation gives approximate bounds on values of
time variable $\tau$ allowed by the restriction on G-dot.

It should be stressed that the solution under consideration with
$P < 0$, $d_1 = 1$ and $d_5 = 3$ takes place when the
configuration of branes, the matrix $(h_{\alpha\beta})$ and
dilatonic coupling vectors $\lambda_a$, obey the relations
(\ref{1.7}), (\ref{1.10}) with $K < 0$. This is not possible when
$(h_{\alpha\beta})$ is positive definite, since in this case $K >
0$. In the next section we give an example of a setup obeying
(\ref{1.7}) and (\ref{1.10}), by introducing "phantom" fields.

\subsection{Example}

Let us consider the following particular example: $N_1 = N_2$,
i.e. the ranks of forms are equal, and $l = 2$, $(h_{\alpha\beta})
= - (\delta_{\alpha\beta})$, i.e. there are two "phantom" scalar
fields. We also put $d_3 = d_4$.

Then relations (\ref{1.7}) and(\ref{1.10}) read as
\beq
\label{3.1nn} \vec{\lambda}_1^2 = \vec{\lambda}_2^2 =d(I) \bigl(
1+ \frac{d(I)}{2-D}\bigr) - K, \eeq and \beq \label{3.2nn}
\vec{\lambda}_1 \vec{\lambda}_2 =d_{\cap} + \frac{(d(I))^2}{2 - D}
+ \frac{1}{2} K, \eeq where $d(I) = d(I_1) = d(I_2) = 1 + d_2 +
d_3$, $d_{\cap} = d(I_1 \cap I_2) = 1 +d_2$ and $K < 0$. Relations
(\ref{3.1nn}) and (\ref{3.2nn}) are compatible since it may be
verified that they imply
 \beq \label{3.3}
 \frac{\vec{\lambda}_1 \vec{\lambda}_2}{|\vec{\lambda}_1| |\vec{\lambda}_2|}
 \in (-1, +1),
 \eeq
i.e. vectors $\vec{\lambda}_1$, $\vec{\lambda}_2$ belonging to
Euclidean space $\R^2$ and obeying relations (\ref{3.1nn}),
(\ref{3.2nn}) do exist. The left side of (\ref{3.3}) gives $\cos
\theta$, where $\theta$ is the angle between these two vectors.

We get in this special case
 \beq \label{3.4}
 A = \frac{4 d(I)}{K (D-2)}.
 \eeq
For $K \to - \infty$ the allowed time interval $(\tau_{-},\tau_{+})$ of
accelerated expansion obeying G-dot restriction (\ref{2.10}) vanishes, i.e.
$\tau_{+} - \tau_{-} \to 0$ (see (\ref{2.13n})).

\section{Conclusions}

In this paper we considered different cosmological models in
diverse dimensions   leading to relatively small time variation of
the effective gravitational constant $G$.

We estimated the possible variations of the gravitational constant
G in the framework of a generalized (Bergmann-Wagoner-Nordtvedt)
scalar-tensor theory of gravity on the basis of the field
equations, without using their special solutions. Specific
estimates were essentially related to the values of other
cosmological parameters (Hubble and acceleration parameters, dark
matter density etc.), but the values of $\dot{G}/G$ compatible
with modern observations did not exceed $10^{-12}$.

We considered also the multidimensional cosmological model with an
$m$-component anisotropic (``perfect'') fluid.  The
multidimensional Hilbert-Einstein equations led to relations
between $\dot G$ and cosmological parameters.

In case of two factor spaces with non-zero curvatures without
matter, we have suggested a mechanism for predicting small $\dot
G$. When the 3-space has a negative curvature and the internal
space has a positive curvature, we got at some time interval an
accelerating expansion of our 3-dimensional space and a small
value of $\dot{G}/G$. We have shown that this result is compatible
with the exact $1+ 3 + 6$ solution  from \cite{GIM}. (Recall that
only three exact solutions are known for a vacuum cosmological
model with a product of two Einstein spaces, see \cite{GIM}).

We also presented another example where two factor spaces are
Ricci-flat and for a  two-component example (dust + 5-brane) we
obtained a small enough variation of $G$.

Besides, we considered multidimensional cosmological models
describing the
 dynamics of $n+1$  Ricci-flat factor spaces $M_i$ in the presence of a
 one-component anisotropic (perfect) fluid with pressures in all spaces
 proportional to the density: $p_{i} = w_i \rho$, $i = 0,...,n$. Solutions
 with accelerated expansion of our 3-dimensional space $M_0$ and small
 enough variation of the gravitational constant $G$ were found.
These solutions have either exponential  or power-law behavior of
scale factors w.r.t. synchronous time variable. In both cases they
exist for two
 branches of parameter $w_0$.  The first branch describes superstiff
 matter with $w_0 > 1$ (and negative energy density) ,
 while the second one may contain phantom matter with
 $w_0 < - 1$ (and positive energy density).
 Here, contrary to the two-curvature model,
 the experimental bounds on $\dot{G}$ are satisfied for all allowed
 values of the synchronous time variable.

 We considered an $S$-brane solution with two
non-composite electric branes and a set of $l$ scalar fields as
well. The solution, corresponding to Lie algebra $A_2$, contains
five factor spaces, and the fifth one, $M_5$, is interpreted as
our 3D space. As in the model with two non-zero curvatures, we
found that there exists a time interval where accelerated
expansion of our 3-dimensional space co-exists with a small enough
value of $\dot{G}/G$ obeying the experimental bounds. Similar
results for other rank 2 algebras were obtained in \cite{IKM-08}.

Thus, here we have shown that there exist different possible ways
of explanation of relatively small time variation of   the
effective gravitational constant $G$ compatible with modern
cosmological data (e.g. acceleration): we may consider either
4-dimensional scalar-tensor theories or multidimensional
cosmological models with different matter sources. The
experimental bounds on $\dot{G}$ may be satisfied either in some
time interval or for all allowed values of the synchronous time
variable (from $(0, + \infty)$ for power-law case or from $(-
\infty, + \infty)$ for the exponential case.

We considered recently \cite{BKMR} also the multidimensional
gravity with a Lagrangian containing the Ricci tensor squared and
the Kretschmann invariant. In a Kaluza-Klein approach with a
single compact extra space of arbitrary dimension, with the aid of
a slow-change approximation (as compared with the Planck scale),
we built a class of spatially flat cosmological models in which
both the observed scale factor $a(\tau)$ and the extra-dimensional
one, $b(\tau)$, grow exponentially at large times, but $b(\tau)$
grows slowly enough to admit variations of the effective
gravitational constant $G$ within observational limits.



\end{document}